\documentclass[a4paper,11pt]{article}
\usepackage{pos}
\usepackage{graphicx}

\title{Effective Field Theories and Lattice for Hard Probes}

\author*[a]{Nora Brambilla}

\affiliation[a]{Physik Department, Technische Universit\"at M\"unchen,\\
James-Franck-Strasse 1, 85748 Garching, Germany}

\emailAdd{nora.brambilla@ph.tum.de}

\abstract{ Effective Quantum Field Theories  and QCD Lattice
  methods have become more and more complementary
  and mutually supportive  in the study of Hard Probes. I present  some of the progress that this
alliance already delivered and I discuss future opportunities.}

\FullConference{%
  HardProbes2020\\
  1-6 June 2020\\
  Austin, Texas}

\begin{document}
\maketitle

\section{Introduction}

Effective Field Theories (EFTs) and QCD Lattice methods have become more and more complementary
and mutually supportive  in the study of Hard Probes.
Since the physical problem is tremendously complex and should be addressed in the underlying field
theory which is QCD, the combination of EFTs and lattice has allowed us to make
big progress in the last times. On one hand, the EFT is pulling out scales from observables
in a controlled way, separating them.
On the other hand, the low energy contributions that the EFT has factorized can be evaluated on the lattice.
This  greatly reduces the complexity of the problem allowing the lattice to target the nonperturbative part
directly, appropriately defined in the EFT in terms of gauge invariant, purely gluonic  objects, a big simplification with
respect to an ab initio lattice calculation of an observable. This last is much more difficult because 
still contains all the physical scales of the problem.
In this framework  also quenched lattice calculations can be pretty useful because at least at higher energy 
the flavor dependence is accounted for by the EFT  matching coefficients.
In relation to hard probes  it has payed off  to combine different EFTs, those who describe the 
hot medium  properties like Hard Thermal Loop  (HTL)  in real time or Electrostatic  QCD (EQCD) and  
Magnetostatic QCD (MQCD) in euclidean time and 
 those describing  processes in the medium like for example Soft Collinear Effective Field Theory (SCET) 
 for jets and nonrelativistic EFTs like Non Relativistic QCD (NRQCD)  or potential Nonrelativistic QCD
 (pNRQCD) for heavy quarks and quarkonium.
In particular,  EFTs allow to resum infinite class of diagrams related to the existence of a physical
scale, to give a precise  quantum field theoretical definition to objects of great phenomenological interest like
the potential or the transport coefficients, to factorize contributions of different scales, to
give a systematical framework for calculation of observables inside field theory.
In this talk I will focus on  heavy probes and  examples of what discussed  above 
will include the free energies,  the heavy quark potential in medium, the heavy quark transport coefficients 
and the nonequilibrium evolution of quarkonium in the fireball. 

To boost the  EFT and lattice interface we founded the TUMQCD lattice collaboration
\cite{tumqcd,Bazavov:2016uvm,Bazavov:2018wmo,Brambilla:2020siz}. In this talk I will review
results at the interface of EFT and lattice.

\section{Heavy probes: Heavy Quarks and Quarkonium}
Heavy quarks  are produced at the beginning of the collision  and remain up to the end.
The heavy-quark mass  $m$ introduces a large scale,
whose contribution may be factorized and computed in 
perturbation theory ($m \gg \Lambda_{QCD},  \alpha_s(m) \ll 1$). 
Low-energy scales  are sensitive  to the temperature   $T$  and even if
nonperturbative they  may be accessible via lattice calculations.
Quarkonia are special  hard probes
because being nonrelativistic bound states, they are multi-scale systems.
They are endowed with three energy scales, the scale of the quark mass $m$ (hard scale), the scale of the momentum transfer
$p \sim mv$ (soft) and the scale of the binding energy $E\sim mv^2$, being $v$ the
velocity in the bound state and being $mv$ proportional to the reverse of the size $r$ of the system. 
If $v$ is smaller than 1 these scales are separated.
In \cite{Brambilla:2004jw}
it was summarized  how to construct an effective field theory
called potential NonRelativistic QCD (pNRQCD)
that allows to define the potentials both in the case of a
perturbative (weakly coupled pNRQCD)  and of a nonperturbative soft scale (strongly coupled pNRQCD):
they are the matching coefficients of the
EFT and they are well defined in the matching procedure. pNRQCD  has the Schr\"odinger equation
as zero problem and it is constructed to be equivalent to QCD order by order in the expansion.
In the case of strongly coupled pNRQCD the potentials are given in terms of generalized Wilson loops
to be calculated on the lattice. In weakly coupled pNRQCD the degrees of freedom are color singlet and color octet
quark-antiquark pair (together with US gluons), in strongly coupled pNRQCD we have only color singlets.

\subsection{pNRQCD at finite T}

Considering quarkonium in  the hot QCD medium also the thermal  scales of the Quark Gluon Plasma  (QGP) are emerging:
the temperature $T$, the Debye mass $m_D \sim g T$ related to the (chromo) electric screening and the scale  $g^2 T$
related to the  (chromo)magnetic screening. In a weakly coupled plasma the scales are separated and
hierarchically ordered.
In a series of papers \cite{Brambilla:2008cx,Brambilla:2010vq,Biondini:2017qjh,Brambilla:2013dpa,Brambilla:2011sg,Escobedo:2008sy}
a pNRQCD at finite $T$ description has been constructed.

\subsection{The Potential}

pNRQCD at finite $T$  allows us to give a clear and systematic definition of what is 
the quarkonium potential  in medium. This  has been investigated for years
using many phenomenological assumptions,  spanning from the internal energy to the  free energy,  either the average free energy or the singlet one
which is gauge dependent.
 The EFT gives us for the first time the possibility to define what is this potential:
 it is the matching coefficient   of the EFT that results from the integration of all the scales above the scale of the binding energy 
 and it is the object that has to be inserted in the Schr\"odinger  equation, the
 zero order equation in  pNRQCD  describing the real time evolution of the  $Q\bar{Q}$ pair in medium.
 When $T$ is bigger than the energy, the potential depends on the temperature, otherwise not.
 Thermal effects appear in any case in the nonpotential contributions to the energy levels.
 We assumed that the bound state exist for $T\ll m$ and $1/r \ge m_D$, we worked in the weak coupling limit and we consider all possible scales hierarchies
 \cite{Brambilla:2008cx}.
We found that the thermal part of the potential has a real part (roughly described by the free energy) 
and an imaginary part. The imaginary part comes from two effects:
the Landau damping \cite{Laine:2006ns,Escobedo:2008sy,Brambilla:2008cx}, an effect
existing also in QED, and the singlet to octet transition, existing
only in QCD \cite{Brambilla:2008cx}. Which one dominates depends on the ratio  between $m_D$ and $E$.
In the EFT one could show that the imaginary part of the potential related to the Landau damping comes
from inelastic parton scattering \cite{Brambilla:2013dpa} and the singlet to octet transitition from gluon dissociation \cite{Brambilla:2011sg}. 
{ \it  The existence of the imaginary part 
 changed our paradigm for quarkonium suppression
as the state  dissociates  for this reason well before that the conventional screening becomes active  
\cite{Laine:2006ns,Escobedo:2008sy,Brambilla:2008cx}.
These large $T$ dependent imaginary parts  call for an appropriate framework to describe the nonequilibrium evolution
of quarkonium in medium: the open quantum system framework as we will discuss in Sec.4.}

The pattern of thermal corrections is pretty interesting  \cite{Brambilla:2008cx}: when $T < E$ thermal corrections are
only in the energy; for $T> 1/r, 1/r>m_D$ or $1/r >T>E$ there is no exponential screening and $T$ dependent power like
corrections appear; if  $T> 1/r, 1/r\sim m_D$ we have exponential screening but the imaginary part of the static potential is
already bigger than the real one and dissociation already happened.
Once the potential has been calculated, the EFT gives the systematic framework to obtain the thermal energies: in \cite{Brambilla:2010vq}
it was performed the first QCD calculation of the thermal contributions to the  $\Upsilon(1S)$  mass and width
at order $m \alpha_s^5$ at LHC below the dissociation temperature
of about 500 MeV. This calculation is very important because it gives the parametric $T$ dependence of this observables. The width goes linear
in $T$ in the dominant term and this has been confirmed by lattice calculations of the spectrum \cite{Aarts:2011sm}.
These findings in the EFT in perturbation theory have inspired many subsequent nonperturbative calculations of the static potential
at finite $T$. In particular the Wilson loop at finite $T$ has been calculated on the lattice
\cite{Rothkopf:2011db,Rothkopf:2019ipj}
finding hints of a large imaginary parts.
These calculations are pretty challenging and refining of the extraction methods are currently in elaboration
\cite{Petreczky:2018xuh,Bala:2019cqu}.

\section{Free Energies and  Polyakov Loops}
Free energies and Polyakov loop calculations have been always very prominent in QCD at finite $T,$ see e.g. the reviews
\cite{Bazavov:2020teh,Ghiglieri:2020dpq}
Here, I report some recent developments at the interface of EFTs and lattice. We  calculated the Polyakov
loop and the Polyakov loop correlators both in perturbation theory and using EFTs to resum scales contributions
in   \cite{Berwein:2015ayt,Berwein:2013xza,Berwein:2017thy,Brambilla:2010xn}
and on the lattice to obtain these quantities fully nonperturbatively in \cite{Bazavov:2016uvm,Bazavov:2018wmo}.
In particular:  the Polyakov loop has been computed up to order $g^6$,
the (subtracted) $Q\bar{Q}$ free energy has been computed at short distances up to
corrections of order $g^7(rT)^4$, $g^8$, the (subtracted)  $Q\bar{Q}$
free energy has been computed at screening distances up to
corrections of order $g^8$; the singlet free energy has been computed at short distances up to corrections of
order $g^4(rT)^5$, $g^6$; the singlet free energy has been computed at screening distances up to
corrections of order $g^5$ \cite{Berwein:2015ayt,Berwein:2017thy,Brambilla:2010xn}.

From the lattice simulations and from comparison to the perturbative results we
could obtain the following important outcomes \cite{Bazavov:2016uvm,Bazavov:2018wmo}: lattice calculations
are consistent with weak-coupling expectations in the regime of application
of the weakly coupled resummed perturbation theory which confirms the predictive power of the EFT;
the crossover temperature to the quark-gluon plasma is $153+ 6.5- 5$ MeV  as 
extracted from the entropy of the Polyakov loop;  the screening sets in at $rT \sim 0.3-0.4$ (observable dependent),
consistent with a screening length of about $1/m_D$; asymptotic screening masses are
about $2m_D$ (observable dependent); the first determination of the color octet $Q\bar{Q}$
free energy has been obtained \cite{Bazavov:2018wmo} further investigated with a  different setup in
\cite{Bala:2020tdt}.

In the EFT/pNRQCD framework the Polyakov loop correlator $P_c$  can be  decomposed as
$$
P_c(r,T) = \frac{1}{N_c^2}[ e^{-{ f_{s}(r,T,m_D)}/T} + (N_c^2-1)e^{-{f_{o}(r,T,m_D)}/T} 
+ {\cal O} (\alpha_s^3(rT)^4 )]
$$
with
$ f_{s}  =$ $Q\bar{Q}$-color singlet free energy, $ f_{o}  =$ $Q\bar{Q} $-color octet free energy 
to be matched from the singlet and octet pNRQCD propagators
$$
\frac{\langle S({\bf r}, {\bf  0},1/T)S^\dagger({\bf r},{\bf 0},0)\rangle}{\cal N}=
e^{-V_s(r)/T}(1+ \delta_s) \equiv e^{- f_{s}(r,T,m_D)/T}$$
and
$$
\frac{\langle O^a({\bf r}, {\bf 0},1/T)O^{a\,\dagger}({\bf r},{\bf 0},0)\rangle}{\cal  N} = e^{-V_o(r)/T}
[(N_c^2-1) \langle P_A \rangle  +  \delta_o ]
\equiv (N_c^2-1)e^{-f_{o}(r,T,m_D)/T}
$$
where $V_s$ and $V_O$ are the singlet and octet static potentials,
$\delta_s$ and $\delta_o$ stand for thermal loop corrections to the singlet/octet propagators,
${\cal N}$ is a normalization and $\langle P_A \rangle $ is the average value of an adjoint Wilson line.

We may identify two possible regimes:
low temperatures, $T \ll V_s$ (or $rT \ll \alpha_s$),: $P_c \approx \frac{e^{-{V_s/T}}}{N_c^2}$;
high temperatures, $T \gg V_s$ (or $rT \gg \alpha_s$): 
$P_c$ is a linear combination of  $e^{-{f_s/T}}$ and $e^{-{f_o/T}}$.
A strict perturbative expansion in $\alpha_s$ corresponds to this regime.
These regimes have been validated with the lattice calculations  \cite{Bazavov:2018wmo}.

The free energies turn out not to be the objects to be used as a potential
in the Schr\"odinger equation 
even if the singlet  free energy may provide a good approximation of the real part of the static  potential.

Differently from the Polyakov loop and the Polyakov loop correlator, the cyclic Wilson loop
 is divergent after charge and field
 renormalization. This divergence is due to intersection points \cite{Berwein:2013xza}.
 In  \cite{Berwein:2013xza} it has been shown that this produces 
  that the cyclic Wilson loop mixes under
  renormalization with the correlator of two Polyakov loops.The resulting renormalization
  equation has been  tested up to order $g^6$ and used to resum the leading
  logarithms associated with the intersection divergence.
The cyclic Wilson loop free energy has been computed at short distances up to
corrections of order $g5 + LL$ resummation and a renormalization prescription relevant for lattice
evaluation has been given \cite{Berwein:2013xza}.

\section{The quarkonium nonequilibrium evolution in medium: EFTs,
  Open Quantum Systems (OQS) and lattice}

The large imaginary parts appearing in the static potential motivated  us to introducing an appropriate
framework to describe the real time nonequilibrium evolution of quarkonium in the QGP medium.
In \cite{Brambilla:2016wgg,Brambilla:2017zei,Brambilla:2019tpt} we have developed an open quantum
system (OQS) framework (for OQS see \cite{Akamatsu:2020ypb} for a review and 
the seminal paper \cite{Akamatsu:2014qsa})
rooted in pNRQCD at finite $T$ that is fully quantum, conserves the
number of heavy quarks and consider both color singlet and color octet quarkonium degrees of freedom.
This has also been reported by the talk of Miguel Escobedo at this conference \cite{miguel}.

We distinguish the environment (QGP) characterized by the scale $T$ and the system (quarkonium) characterized
by the scale $E$.
We identify the inverse of E with the intrinsic time scale of the system: $\tau_S\sim 1/E$
and the inverse of $\pi T$ with the correlation time of the
environment: $\tau_E\sim 1/(\pi T)$. If the medium is in thermal equilibrium,
or locally in thermal equilibrium, we may understand $T$ as a temperature, otherwise is
just a parameter. The medium can be strongly coupled.
The evolution of the system is characterized by a relaxation time $\tau_R$   that  is estimated by the inverse
of the color singlet self-energy diagram in pNRQCD at finite $T$.
We select quarkonia states with a  small radius (Coulombic) for which $1/r \gg \pi T, \Lambda_{QCD}$ and we
consider $\pi T \gg E$.

In this framework ,in~\cite{Brambilla:2017zei}, a  set of  master equations  governing the
time  evolution of  heavy quarkonium  in a  medium were  derived.
The
equations  follow  from  assuming  the  inverse  Bohr  radius  of  the
quarkonium  to  be  greater  than  the energy  scale  of  the  medium,  and model  the  quarkonium as  evolving in  the
vacuum up  to a time  $t=t_{0}$ at  which point interactions  with the
medium begin.
The equations express the time evolution of the density
matrices  of the  heavy quark-antiquark  color singlet,  $\rho_s$, and
octet  states, $\rho_o$,  in  terms  of the  color  singlet and  octet
Hamiltonians, $h_s  = {\bf p}^2/M  - C_F\alpha_s/r  + ...$ and  $h_o =
{\bf p}^2/M + \alpha_s/(2N_cr) +  ...$, and interaction terms with the
medium, which, at order $r^2$  in the multipole expansion, are encoded
in  the self-energy  diagrams of the EFTs.
 These
interactions account for  the mass shift of  the heavy $Q\bar{Q}$
pair induced by the medium, its decay width induced by the medium, the
generation    of   $Q\bar{Q}$ color   singlet    states   from
   $Q\bar{Q}$  color octet states interacting with the medium and the
generation of   $Q\bar{Q}$  color  octet states from $Q\bar{Q}$
(color singlet or octet) states interacting with the medium.
The
leading order  interaction between  a heavy  $Q\bar{Q}$
  field and
the  medium   is  encoded  in   pNRQCD  in  a   chromoelectric  dipole
interaction, which  appears at  order $r/m^0$  in the  EFT  Lagrangian.
The approach gives us  master equations, in general non Markovian,
for the out of equilibrium evolution of the color singlet and color octet matrix densities.
The system is in non-equilibrium because through interaction with the environment
(quark gluon plasma) singlet and octet quark-antiquark states continuously transform in
each other although the total number of heavy quarks is conserved.

Assuming that any energy scale in the medium is
larger than the heavy   $Q\bar{Q}$
 binding energy $E$, in particular that
$\tau_R \gg \tau_E$, we obtain a Markovian evolution while  the chosen hierarchy of scales
implies $\tau_s\gg \tau_E$ qualifying the regime as quantum Brownian motion.
In this situation we can reduce the general master equation to a Linblad form:
\begin{equation}
	\frac{\mathrm{d}\rho}{\mathrm{d}t}=-i[H,\rho]+\sum_{n}\left( C_{n}\rho C_{n}^{\dagger}-\frac{1}{2}\left\{ C_{n}^{\dagger}C_{n},\rho \right\} \right),
\label{eq:lindblad}
\end{equation}
where $H$ is a Hermitian operator, and $C_{n}$ are known as {\em collapse operators}. 
These operators were computed in~\cite{Brambilla:2016wgg,Brambilla:2017zei}. 
If we assume an isotropic medium and the quarkonium at rest with respect to the medium, in the large time limit
the equation \ref{eq:lindblad}
 assumes a particular simple form:
\begin{eqnarray}
&& \rho=\left(\begin{array}{c c}
\rho_s & 0 \\
0 & \rho_o
\end{array}
\right) \\
&& H = \left(\begin{array}{c c}
{ h_s} & 0\\
0 & { h_o}
\end{array}\right)
+ \frac{r^2}{2}\,{ \gamma(t)}\, 
\left(\begin{array}{c c}
1 & 0\\
0 & \frac{7}{16}
\end{array}\right)\,,
\\
&& C^0_i=\sqrt{\frac{ \kappa(t)}{8}}\,r^i\left(\begin{array}{c c}
0 & 1\\
\sqrt{8} & 0
\end{array}\right), \qquad
C^1_i=\sqrt{\frac{5{ \kappa(t)}}{16}}\,r^i\left(\begin{array}{c c}
0 & 0\\
0 & 1
\end{array}\right)
\end{eqnarray}

Interestingly enough in this case the properties of the QGP are encoded in two
transport coefficients: the heavy quark momentum diffusion coefficient, $\kappa$, and its  dispersive counterpart
$\gamma$ which are given by time integrals of appropriate gauge invariant correlators at finite $T$ given 
by the integral of gauge invariant correlators of chromoelectric fields:
\begin{alignat}{3}
	\kappa&=&&\frac{g^{2}}{6N_{c}}\int^{\infty}_{0}\mathrm{d}s~\Big \langle \left\{ E^{a,i}(s,{\bf 0}),E^{a,i}(0,{\bf 0}) \right\} \Big \rangle, \label{eq:kappa_def} \\
	\gamma&=-i&&\frac{g^{2}}{6N_{c}}\int^{\infty}_{0}\mathrm{d}s~\Big \langle \left[ E^{a,i}(s,{\bf 0}),E^{a,i}(0,{\bf 0}) \right] \Big \rangle. \label{eq:gamma_def}
\end{alignat}
In the case of a nonperturbative QGP, these objects are nonperturbative and should be evaluated on the lattice.
Once the Linblad equation is solved and evolved up to freeze out one could obtain 
observables like the $R_{AA}$ and the $v_2$ by projecting over the quarkonium states of interest and compare to the experimental data at LHC, see 
\cite{Brambilla:2016wgg,Brambilla:2017zei}. 

{\it Notice that  in this case the OQS/pNRQCD framework allows us to use input from a lattice QCD in
equilibrium calculation to describe the nonequilibrium evolution of quarkonium in medium.}
I
n \cite{Escobedo:2020tuc} using the static limit the evolution equations have been obtained for quarkonia
of any radius. In \cite{inprep} we are devoloping  a new computational approach to solve the Linblad equations
in a more efficient way that allows to couple the quarkonium system to the full hydrodynamical evolution.

The semiclassical limit of similar equations have been studied in 
\cite{Blaizot:2017ypk} and the relevance of correlated versus noncorrelated noise in
\cite{Sharma:2019xum}. In \cite{Yao:2020kqy,Yao:2018nmy} using this same pNRQCD and OQS framework and a particular
scales hierarchy,  transport equations have been obtained for the study of quarkonium in medium,
in particular a semiclassical Boltzmann equation has been obtained and in the case of the
differential  reaction  rate,  the  information on the QGP is contained
in a novel   chromoelectric  gluon correlator involving also  staple-shaped Wilson lines, in a way
similar   to  what happens  at $T=0$ in the gluon parton distribution functions,
the gluon transverse momentum dependent parton distribution functions 
and in  the quarkonium production cross section factorization  in pNRQCD  \cite{Brambilla:2020ojz}.

\section{Transport Coefficients $\kappa $ and $\gamma $}

\begin{figure}[!tbp]
  \centering
  \begin{minipage}[b]{0.4\textwidth}
    \includegraphics[width=\textwidth]{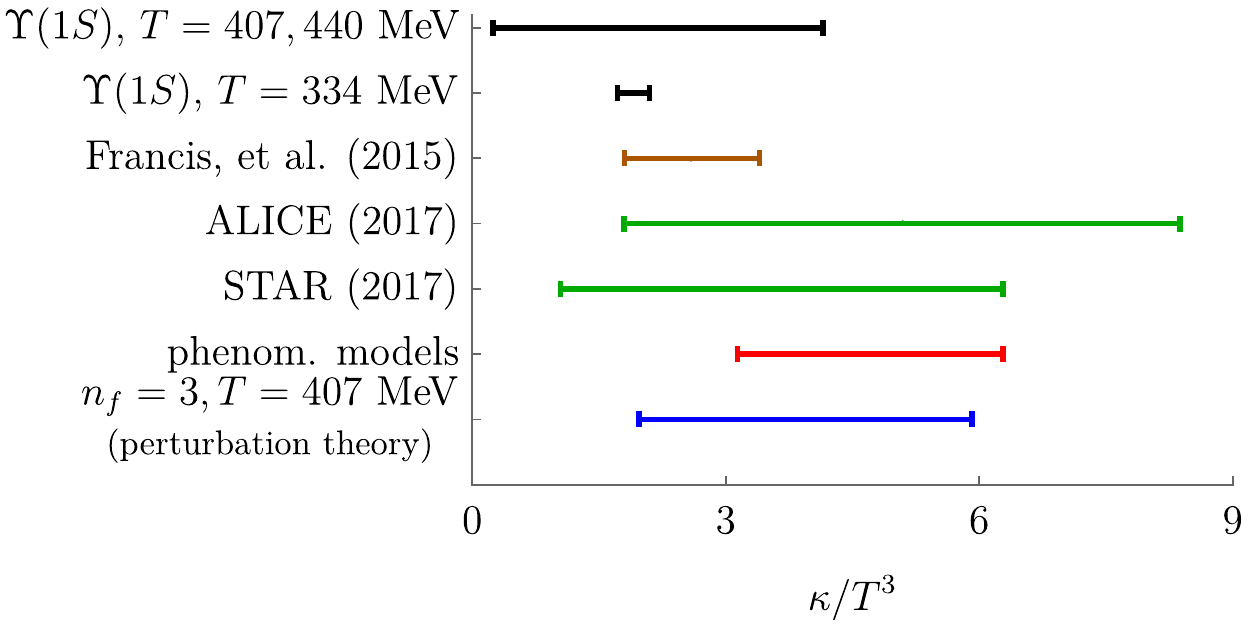}
    \caption{Extraction of $\kappa/T^3$, see the text for details.}
    \label{fig1}
  \end{minipage}
  \hfill
  \begin{minipage}[b]{0.4\textwidth}
    \includegraphics[width=\textwidth]{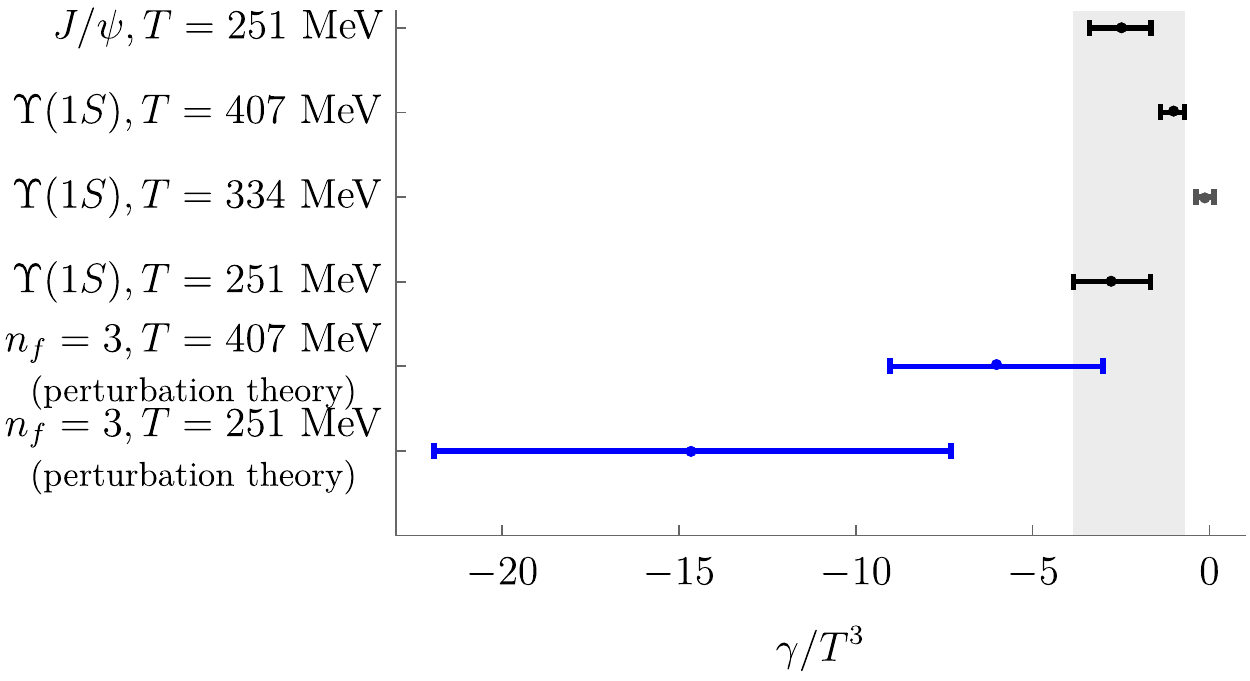}
    \caption{ Extractions of  $\gamma/T^3$, for details see the text.
    }
    \label{fig2}
  \end{minipage}
\end{figure}

The heavy quark momentum diffusion coefficient, $\kappa$, is an object of special interest in the literature,
  but one which has proven notoriously difficult to estimate,
  despite the fact that it has been computed by weak-coupling methods at next-to-leading order accuracy, 
  and by lattice simulations of the pure SU(3) gauge theory.
  Another coefficient, $\gamma$, has been recently identified in the OQS description of quarkonium nonequilibrium
  evolution.   It can be understood as the dispersive counterpart of $\kappa$.
  Nothing is known about $\gamma$, apart from its leading order, weak-coupling expression
  \cite{Brambilla:2008cx,Eller:2019spw}.
  Both $\kappa$ and $\gamma$ are, however, of foremost importance in heavy quarkonium physics
  as they entirely determine the in and out of equilibrium dynamics of quarkonium in a medium, 
  if the evolution of the density matrix is Markovian, and the motion, quantum Brownian.
  The EFT allows to relate such coefficients to  quarkonium thermal energy shifts and widths.
  Precisely  \cite{Brambilla:2019tpt},
  using quarkonia with a small radius (Coulombic) to probe the strongly coupled QGP, we get the relations
  \begin{align}
	\Gamma &= 3 <r>^{2}\kappa, \label{eq:kappa} \\
    \delta M &= \frac{3}{2} <r>^{2}\gamma \label{eq:gamma},               
\end{align}
where $\Gamma$ and 	$\delta M$ are the thermal width and mass shift and $<r>$ is the average radius of the given
quarkonium state. Then, using 2+1 flavors  lattice calculations of $\Gamma$ and $\delta M$
\cite{Aarts:2011sm,Kim:2018yhk}  we could obtain the unquenched values for $\kappa$ and $\gamma$ shown in
Figs. \ref{fig1} and  \ref{fig2}.
In Fig.\ref{fig1},  the first entry (black bar) shows $\kappa/T^3$ as obtained from Eq. (\ref{eq:kappa})
using the lattice data of Refs.~\cite{Aarts:2011sm,Kim:2018yhk} 
for the lower and upper bounds of the thermal decay width of the $\Upsilon(1S)$.
The second entry (brown bar) reports the (quenched) lattice estimate of Ref.~\cite{Francis:2015daa}.
The third and fourth entries (green bars) are the determinations based on the ALICE~\cite{Acharya:2017qps}
and STAR~\cite{Adamczyk:2017xur} measurements of the $D$-meson azimuthal anisotropy coefficient $v_{2}$, respectively.
The fifth entry (blue bar) is the leading order (LO) perturbative result with the strong coupling computed 
at the scale  $\pi \times (407\text{ MeV}) = 1.28\text{ GeV}$. We assign to it a 50\% uncertainty.

In Fig. \ref{fig2}
the first three entries (black bars) show $\gamma/T^3$ as obtained from Eq.~(\eqref{eq:gamma}) using lattice data of Ref.~\cite{Kim:2018yhk} 
for the thermal mass shift of the $J/\psi$ and of the $\Upsilon(1S)$ at two different temperatures. The error bars account for the lattice uncertainties only.
The last two entries (blue bars) provide  $\gamma/T^3$ in perturbation theory at leading order at two different temperatures. 
We assign a 50\% uncertainty to these results. The gray band gives our final range.
The resulting range for $\kappa$ is consistent with the earlier determinations,
the one for $\gamma$ is the first non-perturbative determination of this quantity.

{\it This is a clear example of how one could obtain results beyond the present state of the art
(in this case an unquenched lattice calculation of $\kappa$ and $\gamma$) taking advantage of
the alliace between EFT and lattice}.

\subsection{Direct lattice calculation of $\kappa$}

In \cite{Brambilla:2020siz}
we computed the heavy quark momentum diffusion coefficient directly from the correlator
of two chromo-electric fields attached to a Polyakov loop in pure SU(3) gauge theory.

Notice that in general the  lattice calculations of  the transport
coefficients   are  very   difficult.   In fact,  to   obtain  the   transport
coefficients  one has  to reconstruct  the spectral functions  from the
appropriate Euclidean  time correlation functions.  At  low energies, $\omega$,
the  spectral function  has a  peak,  called  the transport  peak,  and
the width    of   the    transport   peak    defines   the    transport
coefficient. Thus, one needs a  reliable determination of the width of
the transport peak in order  to obtain the transport coefficient from
lattice QCD calculations, which is difficult.  In the case of
heavy quarks,  this is even more  challenging because the width  of the
transport peak   is  inversely   proportional   to   the  heavy   quark
mass. Moreover, Euclidean  time correlators are rather  insensitive to
small widths.
{\it The above difficulty  in the determination of  the heavy quark diffusion
coefficient  can be  circumvented  by using EFTs.
Namely, by  integrating out the heavy quark  fields one can
relate the heavy quark  diffusion coefficient to the  correlator of the
chromoelectric    field    strength, as we have previously discussed.     The    corresponding
spectral function does not have a transport peak and the small $\omega$ behavior
is  smoothly connected  to  the UV  behavior  of the spectral  function
The heavy quark diffusion coefficient is given by the intercept
of the spectral function at $\omega= 0$ and no determination of the width of the
transport peak is needed.}. Recently the subleading correction (in the mass expansion) to
$\kappa$ has been calculated and found to be proportional to a correlator of magnetic fields
\cite{Bouttefeux:2020ycy}.

\begin{figure}[!ht]
  \includegraphics[width=8.6cm]{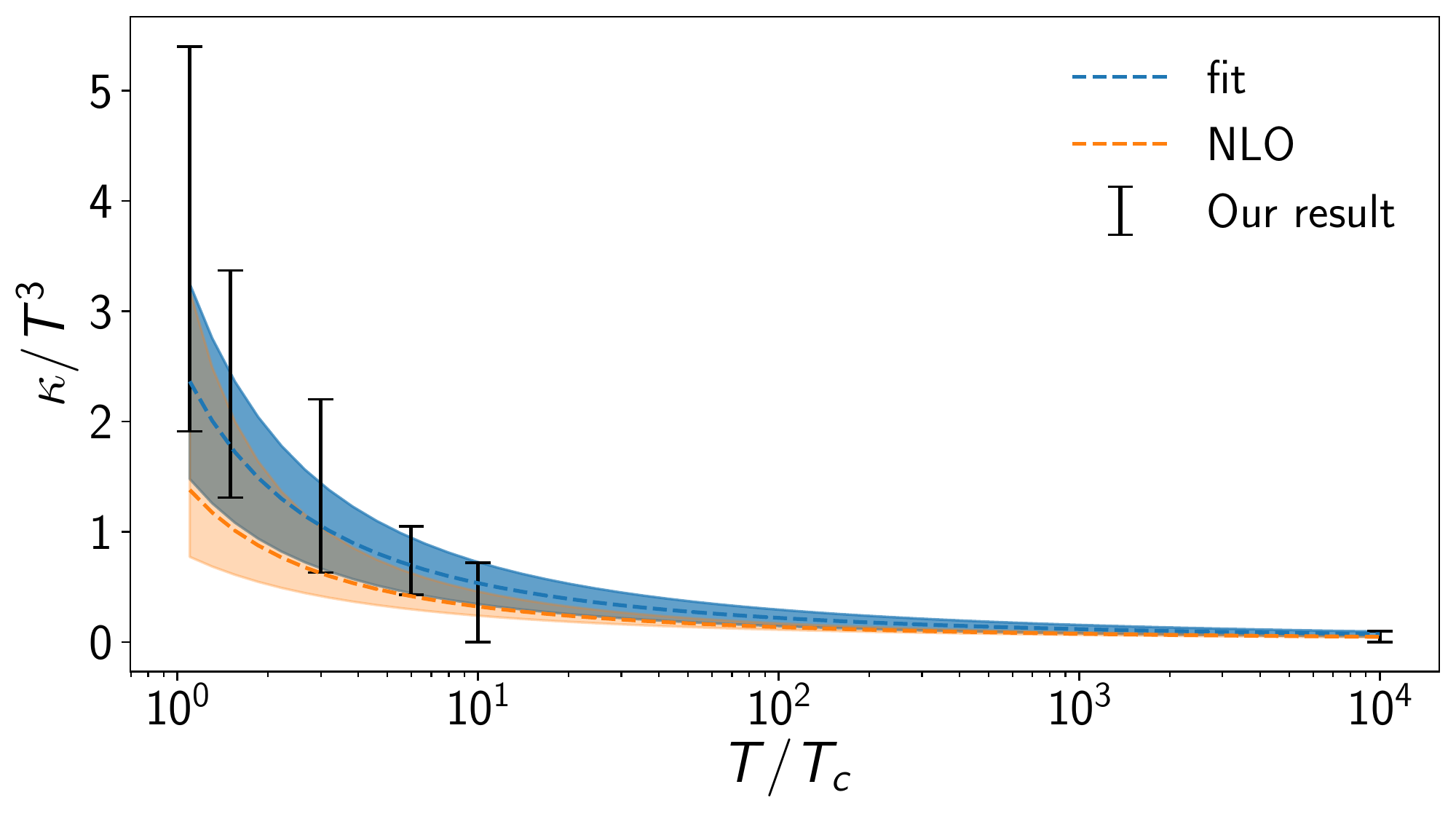}
  \caption[b]{Temperature dependence of our results compared to the NLO result. 
              The shaded bands include the errors coming from varying the scale by a factor 2.
              The blue band also includes the statistical error.
  }
  \label{fig:fitkappac} 
\end{figure}

Using a multilevel algorithm and tree-level improvement,
we studied the behavior of the diffusion coefficient as a function
of temperature in an unprecedented  wide range  of temperature $1.1<T/T_c<10^4$.
At high T is possible 
to compare with the perturbative expansions in the EFT and we find that 
 within errors the lattice results are remarkably compatible with the next-to-leading order
perturbative result, as you see from  Fig. \ref{fig:fitkappac}. 

These results expose for the first time the temperature dependence of $\kappa$ in a large of window of temperature 
and have a great impact on  $R_{AA} $ and $v_2$ predictions and it has been shown solving the Linblad equations
of   \cite{Brambilla:2016wgg,Brambilla:2017zei}    with a $T$ dependent $\kappa$, cf. \cite{inprep}.

This correlator has been recently computed quenched on the lattice  at $T=1.5 T_c$  using the gradient flow in 
\cite{Altenkort:2020fgs}.

\section{Outlook}
We focused on heavy quarks but there are other hard probes  to which the EFT/lattice approach can  be applied.
A  clear example is the calculation  of  $\hat{q}$ that controls jet quenching.
A systematic treatment of a complex phenomenon like jet quenching is possible in an EFT framework owing to the hierarchy of scales that characterizes the system.
These are typically SCET scales  $Q, Q \lambda, Q \lambda^2$ with $\lambda =T/Q$ which characterize the propagation of a very energetic 
parton in the medium and the thermal scales that characterize the medium itself. $\hat{q}$ is the jet quenching parameter, i.e. the mean square transverse momentum 
picked up by the hard parton per unit distance traveled. Using EFT methods $\hat{q}$ can be expressed in terms of gauge invariant gauge field correlators 
to be calculated  on the lattice, see \cite{Benzke:2012sz,Kumar:2020pdl}.
Recently an open quantum system description of jet quenching rooted in the EFT/lattice has been developed \cite{Vaidya:2020cyi}
in analogy to what I have described above for the nonequilibrium evolution of quarkonium in medium.

{\it In summary, the alliance of EFTs, resummed perturbative and lattice methods allows to study hard probes directly in the realm of QCD in a systematic way.
In this framework hard probes become a unique laboratory for the study of the QGP.}
 Some previous disagreements between perturbation theory and lattice appear to be solved.
The EFT factorization at lower energy increase our predictivity power: nonperturbative objects
depends only on the lower energy scale, are reduced in number and formulated such that they can be evaluated on the lattice directly.
EFTs allow us to give the appropriate definition and define a calculational scheme for quantities of huge phenomenological interest.
 EFTs allow us to enlarge the applicability region of lattice see e.g. the non equilibrium evolution of quarkonium in the fireball.
 I have reported results on the free energies, the potential, the thermal spectrum, the $R_{AA}$, some diffusion coefficient but more can be studied in the same framework.
 Many of the correlators factored in the EFT have still to be computed on the lattice.
Calculating these objects on the lattice and developing tools to calculate them also unquenched,  like the gradient flow,
as well as efficient techniques to relate to continuum will have a profound impact on the phenomenology of hard probes.
Many EFTs applications have still to be worked out and new field entered  like nonequilibrium studies with open quantum systems.

\end{document}